\documentstyle[twocolumn,aps,epsfig]{revtex}
\newcommand{\be}{\begin{equation}}
\newcommand{\ee}{\end{equation}}
\newcommand{\bea}{\begin{eqnarray}}
\newcommand{\eea}{\end{eqnarray}}

\newcommand{\lton}{\mathrel{\lower.9ex
                  \hbox{$\stackrel{\displaystyle <}{\sim}$}}}

\begin{document}
\title{Effects of Strong Color Fields on Baryon Dynamics}
\author{Sven~Soff$\,^1$, J\o rgen~Randrup$^1$, 
Horst~St\"ocker$^2$, and Nu Xu$^1$}
\address{$^1$Lawrence Berkeley National Laboratory, 
Nuclear Science Division 70-319, 
1 Cyclotron Road, Berkeley, CA 94720, U.S.A.\\
$^2$Institut f\"ur Theoretische Physik, J.W.\ Goethe-Universit\"at, 
60054 Frankfurt am Main, Germany}
\date{\today}
\maketitle   
\begin{abstract}
\mbox{We calculate the antibaryon-to-baryon ratios,} 
$\overline{p}/p, 
\overline{\Lambda}/\Lambda, \overline{\Xi}/\Xi$, and 
$\overline{\Omega}/\Omega$ for 
Au+Au collisions at RHIC ($\sqrt{s}_{NN}=200\,$GeV).
The effects of strong color fields associated with 
an enhanced strangeness and diquark production probability and 
with an effective decrease of formation times
are investigated. 
Antibaryon-to-baryon ratios increase with the 
color field strength.
The ratios also increase with the strangeness content  
$|S|$. 
The netbaryon number at midrapidity considerably increases with 
the color field strength while the netproton number remains 
roughly the same. 
This shows that the enhanced baryon transport involves a  
conversion into the hyperon sector ({\it hyperonization}) which 
can be observed in the 
$(\Lambda-\overline{\Lambda})/(p-\overline{p})$ ratio.   
\end{abstract}
\narrowtext
\vspace*{-0.8cm}
Ultrarelativistic heavy ion collisions provide a unique tool 
to study elementary matter at energy densities so high that  
a phase transition from partonic deconfined matter 
to hadronic matter is predicted by QCD lattice calculations.   
The properties of such a highly excited state depend  
strongly on the initial conditions such as the specific entropy 
density or the thermalization time. 
A quantum number of utmost interest is 
the baryon number which, prior to the collision, 
is non-vanishing only at beam and target rapidities which   
for energies at the Relativistic Heavy Ion Collider (RHIC) 
(center-of-mass energy per nucleon-nucleon collision 
$\sqrt{s}_{NN} \sim 200\,$GeV, Au+Au) 
are separated by 
\begin{displaymath}
\Delta y \sim 11
\end{displaymath} 
units.
Accordingly, one often assumes complete transparency, i.e., vanishing 
netbaryon number and $\overline{B}/B$-ratios identical with one 
at midrapidity. 
Recent data by the STAR \cite{Adler:2001aq}, 
PHOBOS \cite{Back:2001qr}, 
BRAHMS \cite{Bearden:2001kt}, and 
PHENIX \cite{Adcox:2001mf} collaborations, 
however, show that the midrapidity region $y\sim 0$ obtains 
a considerable finite netproton ($p-\overline{p}$) number in the 
final state. Moreover, the antibaryon-to-baryon ratios are not yet 
equal to one. The $\overline{p}/p$-ratio is about 0.7 for the 
most central collisions. 
Furthermore, rapidity spectra seem to exhibit rather a plateau 
than a minimum at midrapidity and large elliptic flow values 
indicate the need for a large amount of momentum equilibration 
at midrapidity providing additional evidence for a non-transparent 
scenario. 

This strong baryon number transport over more than 5 units in rapidity 
is a novel phenomenon and its underlying mechanism is of utmost 
interest \cite{Busza:1983rj,Sorge:pd,Mishustin:2001ib,Danos:2000mv,Frankfurt:1990wn}. 
It also has a strong impact on other 
questions like the (parton) equilibration \cite{Wang:xy}, 
the equilibration in terms of hadronic yields \cite{Braun-Munzinger:1999qy}
and the build up of collective flow \cite{scheid74}. 

In this Letter we study the effects of strong color fields on the 
dynamics of baryons in relativistic heavy ion collisions. 
We will show that strong color fields lead to an enhanced baryon-antibaryon 
pair production and to a strong modification of 
the baryon transport dynamics. This will affect the antibaryon-to-baryon 
ratios as well as the netproton and netbaryon rapidity distributions. 
Moreover, the netbaryon transport to midrapidity includes 
a {\it hyperonization}, i.e., the conversion of ordinary 
nonstrange baryons to baryons carrying strangeness.   

The idea of strong color electric fields, i.e., of an effectively 
increased string tension in a densely populated colored environment of 
highly excited matter has been suggested already earlier. 
The particle formation process in hadronic collisions 
can be viewed as quantum tunneling of quark-antiquark and gluon pairs
in the presence of a background color electric field. 
It is formed between two receding hadrons which are color charged 
by the exchange of soft gluons while colliding. 
In nucleus-nucleus collisions the color charges may be considerably greater 
than in nucleon-nucleon collisions due to the almost simultaneous 
interaction of several participating nucleons \cite{kajantie85}. 
This leads to the formation of strong color electric fields.
With increasing energy of the target and projectile the number and density of
strings grows, so that they start overlapping, forming clusters,
which act as new effective sources for
particle production \cite{biro84,sor92}.
It has been predicted that the multiplicities 
of, for example, strange baryons or antibaryons should be 
strongly enhanced \cite{sor92,gyulassy90,gerland95,Soff:1999et} once the color 
field strength grows. 
The abundances of (multiply) strange (anti)baryons  
in central Pb+Pb collisions at Cern-SPS \cite{and98a}, for example,  
can only be explained within 
the framework of microscopic model calculations \cite{Soff:1999et}
if the elementary production probability of $s\overline{s}$ pairs, which 
is governed in the string models \cite{Andersson:ia} 
by the Schwinger mechanism \cite{schwing51} 
$\sim \exp (- \pi m_q^2/2\kappa)$, 
is considerably enhanced. 
This corresponds either to 
a dramatic enhancement of the string tension $\kappa$ 
(from the default $\sim 1\,$GeV/fm to $3\,$GeV/fm) or to  
quark masses $m_q$ that are reduced from their constituent quark values 
to current quark values as motivated by chiral symmetry 
restoration \cite{Soff:1999et}. 
In a fully analog way, $\overline{p}$ abundances can be explained 
since an enhanced string tension similarly leads to an 
increased production probability of antidiquark-diquark pairs which 
is needed to account for the experimentally observed 
yields \cite{Bleicher:2000gj}.  
However, this argument, providing additional motivation,  
has to be reconsidered if multifusion processes \cite{rapp},   
e.g., $5\pi \rightarrow \overline{B}B$, that are neglected 
in microscopic transport models based on $2\rightarrow n $ scatterings, 
contribute significantly to the baryon pair production.
A variation of the string tension from $\kappa=1\,$GeV/fm to $3\,$GeV/fm 
increases the pair production probability of 
strange quarks (compared to light quarks) 
from $\gamma_s=P(s\overline{s})/P(q\overline{q})=0.37$ to 0.72. 
Similarly, the diquark production probability is enhanced 
from $\gamma_{qq}=P(qq\overline{q}\overline{q})/P(q\overline{q})=0.093$  
to 0.45. In general, heavier flavors or diquarks ($Q$) are suppressed 
according to the Schwinger formula \cite{schwing51} by 
\begin{equation}
\gamma_Q=\frac{P(Q\overline{Q})}{P(q\overline{q})}= \exp\left(
-\frac{\pi (m_Q^2-m_q^2)}{2 \kappa} \right)\,.
\end{equation} 

Additional motivation of an effectively enhanced string tension in 
a densely colored environment is given due to its relation 
to the Regge slope $\alpha'$. Based on a rotational string picture 
the string tension $\kappa$ is related to the Regge slope 
$\alpha'$ by \cite{goddard,Wong:jf}
\begin{equation}
\kappa=\frac{1}{2 \pi  \alpha'}.
\label{alphakappa}
\end{equation}
The empirical value of the Regge slope for baryons is 
$\alpha'\approx 1\,$GeV$^{-2}$ \cite{green87} that yields 
a string tension of approximately $1\,$GeV/fm. 
However, high-energetic processes dominated by Pomeron exchange, 
characterizing 
the multi-gluon exchange processes existent in high-energetic nucleus nucleus 
collisions, are described by a Regge trajectory (Pomeron) 
with a smaller slope of $\alpha_P'\approx 0.4\,$GeV$^{-2}$ 
\cite{veneziano,collins}. 
According to Eq.\ (\ref{alphakappa}) this translates into a 
considerably larger ({\it effective}) string tension $\kappa$.

It has been suggested recently that the magnitude of a 
typical field strength 
at RHIC energies might be as large as $5-12\,$GeV/fm \cite{Magas:2000jx} 
(as a result of collective effects related to quark-gluon-plasma formation). 
This in turn may 
enhance an anti-flow component resulting in a novel wiggle structure 
(that in general can have different origins
\cite{Magas:2000jx,soff95,Csernai:1999nf,snellings,brachmann})  
around midrapidity in semi-central collisions 
\cite{Magas:2000jx}. 

For our investigation of the baryon dynamics in heavy ion collisions at RHIC 
a microscopic transport approach based on the covariant propagation of
constituent quarks and diquarks accompanied by mesonic and baryonic 
degrees of freedom is applied \cite{urqmd}.
It simulates multiple interactions of ingoing and newly produced 
particles, the excitation and fragmentation of color strings 
and the formation and decay of hadronic resonances. 
Standard type fragmentation functions are used to assign 
the individual hadrons of the string break-up 
a longitudinal momentum fraction $x$. 
Leading nucleons obey the distributions 
\begin{displaymath}
f(x)_n=\exp(-(x-b)^2/(2a^2))
\end{displaymath}
with the maximum at $b=0.42$ and the width $a=0.275$. 
For the produced hadrons a Field-Feynman fragmentation 
function \cite{fieldfeyn} is used 
\begin{displaymath}
f(x)_p=(1/3)((1-x)^d c (d+1)+1-c)
\end{displaymath}
with constants 
$c=1$ and $d=2$.
At RHIC energies, subhadronic degrees of freedom are 
of major importance. 
They are treated in the Ultra-relativistic Quantum Molecular Dynamics 
(UrQMD) model \cite{urqmd} via 
the introduction of formation times for hadrons that are produced in the 
fragmentation of strings \cite{Andersson:ia}.
Leading hadrons of the fragmenting strings contain the valence-quarks 
of the original excited hadron.
They are allowed to
interact even during their formation time, with a reduced cross section
defined by the additive quark model, 
thus accounting for the original valence quarks contained in that
hadron \cite{urqmd}. 
Newly produced (di)quarks are not allowed to  
interact until they have coalesced into hadrons. 
Their formation times are inversely proportional to the 
string tension 
\begin{displaymath}
t_f \sim 1/ \kappa.
\end{displaymath} 
The larger the string tension the shorter and short-living are the 
strings with a certain total energy.
The cross sections for leading quarks and diquarks are 
set to the (later to be formed) hadron hadron cross sections
scaled down by quark counting
\begin{displaymath} 
\sigma_{q(\tilde{M})h}=\sigma_{Mh}/2,\,\sigma_{qq(\tilde{B})h}=
2\sigma_{Bh}/3,\, 
{\rm and}\,  \sigma_{q(\tilde{B})h}=\sigma_{Bh}/3, 
\end{displaymath}
where, e.g.,  
$q(\tilde{M})$ represents a quark of an still unformed meson $\tilde{M}$. 
In this way the overall stopping behaviour of protons 
in central S+S collisions at $E_{\rm lab}=200\,A$GeV can be described 
reasonably \cite{Winckelmann:1996vu}.
The importance of the diquark dynamics, i.e., (re)scatterings based on 
quark model cross sections, for the stopping behavior thus becomes 
apparent \cite{Winckelmann:1996vu,Bleicher:2000pu}.  
These secondary scatterings are important for transporting 
baryon number from projectile and target rapidity closer to midrapidity.
The baryon-baryon collision spectrum is 
largely dominated by diquark degrees of freedom \cite{Winckelmann:1996vu}. 
Only for the very first collisions and at the later stage, 
lower energetic collisions 
involve fully formed baryons.  Additional contributions 
arise from (single) quarks carrying baryon number after 
diquark breaking. The diquark breaking mechanism \cite{capella} 
is assigned a probability of ($\sim 10\% $).

Rather exotic mechanisms, e.g., baryon
junctions \cite{Kharzeev:1996sq,Vance:1998vh},  
have been suggested in order to understand the baryon number
transport in high energetic nucleus nucleus collisions. 
In this approach, baryon number is not necessarily carried by the valence 
quarks but may reside in a non-perturbative configuration of gluon fields. 

There are other important model ingredients 
to the baryon dynamics, i.e., in particular to the baryon transport 
over several units in rapidity and the baryon-antibaryon pair production. 
The total and elastic baryon antibaryon cross sections at high 
energies ($>5\,$GeV/c) are parameterized according to 
\begin{displaymath}
\sigma_{\rm tot/el}(p_{\rm lab})=a+b p_{\rm lab}^n+c \ln^2 p_{\rm lab} 
+ d \ln  p_{\rm lab}
\end{displaymath}
with appropriate constants $a,\,b,\,c,\,d,\,$and$\, n$ 
to fit experimental data \cite{urqmd}.
The $\overline{p}p$ annihilation cross section is parameterized as  
\begin{displaymath}
\sigma_{\rm an}^{\overline{p}p}(\sqrt{s})=\sigma_0  (s_0/s) 
[(a^2s_0)/((s-s_0)^2+a^2s_0)+b].
\end{displaymath} 
For the $\overline{p}n$ and the $\overline{B}B$ cross sections 
we assume 
\begin{displaymath}
\sigma^{\overline{p}n}(\sqrt{s})= \sigma^{\overline{p}p}(\sqrt{s})
\end{displaymath}
and 
\begin{displaymath}
\sigma^{\overline{B}B}(\sqrt{s})=\nu_s \sigma^{\overline{p}p}(\sqrt{s})
\end{displaymath}
with $\nu_s=1$ if the baryons are both nonstrange and 
$\nu_s=\sigma^{\overline{B}B}_{\rm aqm}/\sigma^{\overline{p}p}_{\rm aqm}$
for baryons with $s_i \ge 1$ (anti)strange quarks, respectively.  
The energy-independent additive quark model cross section in units of mb, 
(here only used to scale the energy-dependent $\overline{B}B$ cross section), 
follows from 
\begin{displaymath}
\sigma^{\rm aqm}_{\rm tot}=40(2/3)^{(n_1+n_2)}
(1-0.4 s_1/(3-n_1))(1-0.4 s_2/(3-n_2))
\end{displaymath}
with 
$n_i=1$ for mesons and $n_i=0$ for baryons.
For further details of the model we refer to \cite{urqmd}.

The anti-baryon-to-baryon ratios at midrapidity are shown in Figure~1 
as a function of the strangeness content $|S|$, both 
for $\kappa=1\,$GeV/fm as well as for the strong color 
field scenario (SCF, $\kappa=3\,$GeV/fm). 
\begin{figure}[htp]
\vspace*{-1cm}   
\centerline{\hspace{.4cm}\hbox{\epsfig
{figure=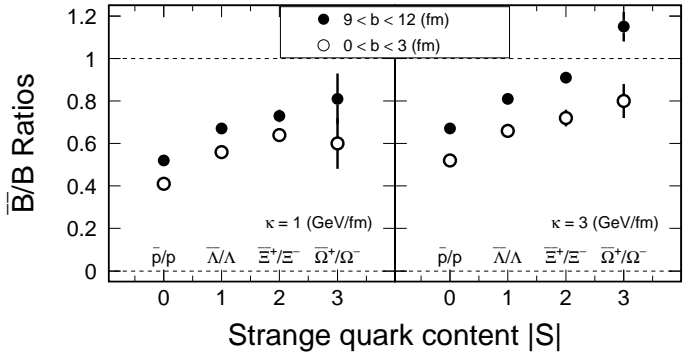,width=10.5cm}}}
\caption{Antibaryon-to-baryon ratios at midrapidity as a function 
of the strangeness content $|S|$ in 
Au+Au collisions at RHIC ($\sqrt{s}_{NN}=200\,$GeV). Calculations with 
a string tension of $\kappa=1\,$GeV/fm are shown on the left 
and the results with strong color fields ($\kappa=3\,$GeV/fm) 
are shown on the right. The open and full circles correspond to  
rather central ($b < 3\,$fm) and peripheral ($9< b <12\,$fm) 
collisions, respectively.}
\label{fig1}
\end{figure}
The following systematics arise from these calculations. 
(i) The $\overline{B}/B$-ratios increase with the strangeness content $|S|$ of 
the baryons. 
(ii) The $\overline{B}/B$-ratios increase with impact parameter $b$. 
There is stronger absorption of antibaryons in central collisions (reducing 
the numerator of the ratio) and more {\it stopping} 
(increasing the denominator).
(iii) Most important, the $\overline{B}/B$-ratios also increase 
with the color field strength $\kappa$.  
The dependence of the $\overline{B}/B$-ratios on $\kappa$ is, a priori, 
not directly transparent. 
On the one hand, an increased pair production 
drives the ratio towards 1. On the other hand, reduced formation times 
and, hence, stronger baryon transport, reduce the ratios.   

In general, the $\overline{B}/B$-ratio is determined through 
the pair production of $\overline{B}B$ pairs, i.e., baryons and antibaryons 
produced in equal amounts, and the collisions which transport 
(net)baryons from beam rapidity to midrapidity. 
Both, the pair production and the number of collisions 
(through smaller formation times) increase with increasing 
field strength $\kappa$. 
So in a simple ansatz one may write 
\begin{displaymath}
\overline{B}/B \sim P(x)/[P(x)+C(x)],
\end{displaymath}
where $P(x)$ is the number of produced $\overline{B}B$ pairs 
and $C(x)$ is the number of (net)baryons (which were transported to 
midrapidity). Both quantities depend on some coupling $x$, 
including the color field strength $\kappa$;  
(other couplings are for example the (re)absorption channels).
If collisions were neglected, i.e., $C(x)=0$, the 
$\overline{B}/B$-ratio equals unity. This case corresponds 
to (freely) decaying strings and yields the oversimplified 
transparency assumption as mentioned in the introduction. 
However, if collisions are taken into account 
(and this is the case in this model where the leading quarks and 
diquark cross sections are always nonzero, even in the formation time,) 
then one obtains $\overline{B}/B$-ratios smaller than unity even 
for small field strengths $\kappa$ for which the 
$qq,\overline{q}\overline{q}$ pair production probability is 
strongly suppressed (see eq.\ (1)). 
For large field strengths $\kappa$, pair production dominates 
resulting in larger ratios $\overline{B}/B \sim P(x)/[P(x)+C(x)] \leq 1$.

In Table~I, the results for the central collisions are compared 
to RHIC data from STAR at 
$\sqrt{s}=130\,$GeV ($10\%$ most central collisions) \cite{vanburen}. 
The results from other collaborations agree within error bars. 
Obviously, the calculations obtained 
with strong color fields agree better with the data. 
Note that feeddown corrections, e.g., protons from $\Lambda$ decays,  
may further reduce the experimentally (uncorrected) ratios. 
Experimental error bars are statistically only.
\begin{table}
\caption{$\overline{p}/p,
\overline{\Lambda}/\Lambda, \overline{\Xi}/\Xi$, and 
$\overline{\Omega}/\Omega$ ratios at midrapidity
for central Au+Au collisions at RHIC for calculations with 
$\kappa=1\,$GeV/fm, with strong color fields and
compared to (partially preliminary) STAR data  
at $\sqrt{s}_{NN}=130\,$GeV.}
\begin{center}
\begin{tabular}{|r||l|l|r|}
\hline
   & $\kappa=1\,$GeV/fm & (SCF) $\kappa=3\,$GeV/fm& exp.\ data \\
\hline
$\overline{p}/p$   & $ 0.41 \pm 0.01 $  &  $0.52    \pm 0.01$    & $0.72\pm 0.01$ \\
$\overline{\Lambda}/\Lambda$   & $0.56  \pm 0.01$ &  $0.66 \pm 0.02   $    &
$0.73 \pm 0.03$ \\
$\overline{\Xi}/\Xi$ & $0.64 \pm 0.03$  & $0.72 \pm 0.04$    & $0.83 \pm 0.03 $\\
$\overline{\Omega}/\Omega$ & $0.60 \pm 0.12$ & $0.80\pm 0.08$ & $0.95 \pm 0.15 $\\
\hline
\end{tabular}
\end{center}
\end{table}

The rapidity spectra for netprotons ($p-\overline{p}$) and 
netbaryons ($B-\overline{B}$) are shown for different centralities 
in Figure~2. 
For the most central collisions, the rather strong maxima at 
$y\approx 3.2$ ($\kappa=1\,$GeV/fm) 
are weakened and shifted closer to midrapidity 
($y\approx 1.8$) in the case of strong color fields ($\kappa=3\,$GeV/fm). 
The number of netprotons at midrapidity seems to be independent 
on the color field strength ($\approx 12$) while the number of 
netbaryons increases (from about $32$ to $42$). 
Hence, the larger netbaryon number at midrapidity results 
from strange baryons. 

\begin{figure}[htp]
\vspace*{-1cm}
\centerline{\hspace{.4cm}\hbox{\epsfig
{figure=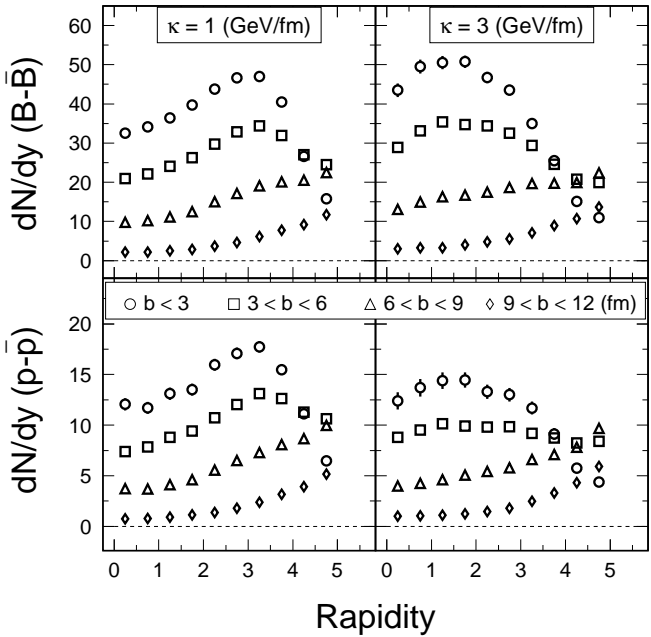,width=10cm}}} 
\caption{Rapidity distributions for net-baryons (top) and 
net-protons (bottom) in 
Au+Au collisions at RHIC ($\sqrt{s}_{NN}=200\,$GeV). Calculations with
a standard string tension ($\kappa=1\,$GeV/fm) are shown on the left 
and the results with strong color fields ($\kappa=3\,$GeV/fm)
are shown on the right. The symbols correspond to different  
centralities ($b < 3\,$fm (circles), $3 < b <6\,$fm (squares),
$6 < b < 9\,$fm (triangles), and $9< b <12\,$fm (diamonds)).  
}
\label{}
\end{figure}

This {\it hyperonization} can also be seen in the 
$(\Lambda-\overline{\Lambda})/(p-\overline{p})$ ratio at midrapidity
which is $\sim 0.45$ 
for calculations with $\kappa=1\,$GeV/fm and strongly enhanced 
$\sim 0.8$ 
for the strong color field (SCF) ($\kappa=3\,$GeV/fm) scenario, corresponding  
to the experimental value \cite{Adcox:2001mf,Adler:2001aq}. 
Note that $\Sigma^0$'s and $\overline{\Sigma^0}$'s are included in the 
$\Lambda$'s and $\overline{\Lambda}$'s, respectively, because 
they cannot be distinguished experimentally.

Another consequence of strong color fields 
is an increase of the pressure (gradients). 
This, in turn, may lead to strongly enhanced 
elliptic flow values as large as observed in the data. This provides  
a possible explanation for the lack between calculations 
that assume the {\it vacuum} string tension 
and that predict elliptic flow values \cite{bleicher2002} 
too small compared to data at RHIC \cite{ackermann2000}.

The transport of other quantum numbers that can be examined  
in AA as well as in pA collisions include strangeness \cite{Soff:1999et} and  
isospin \cite{urqmd,leifels}. 
Those can provide a complementary picture of the baryon number transport 
and do not need to be conserved locally. 

We have calculated the antibaryon-to-baryon ratios, netbaryon and 
netproton rapidity distributions for Au+Au collisions at RHIC energies, and 
have demonstrated the important effects of strong color fields on these 
observables. An enhanced pair production of strange quarks and diquarks 
and reduced formation times 
lead to larger $\overline{B}/B$-ratios and to a stronger 
rapidity loss for the netprotons and netbaryons. More  
netbaryons are transported to midrapidity in the strong color field scenario 
which is accompanied by a hyperonization of these netbaryons. 
This hyperonization can be observed through an enhanced 
$(\Lambda-\overline{\Lambda})/(p-\overline{p})$ ratio and possibly represents 
the clearest signature of the flavor 
equilibration in relativistic heavy ion collisions. 
\vspace*{-0.4cm}
\acknowledgements 
\vspace*{-0.4cm}
We are grateful to M.\ Bleicher, K.\ Schweda, X.\ N.\ Wang
for valuable comments. We thank the
UrQMD collaboration for permission to use the UrQMD transport model.
S.S.\ has been supported by the Alexander von Humboldt Foundation 
through a Feodor Lynen Fellowship.
This research used resources of the National Energy Research Scientific 
Computing Center (NERSC). 
This work is supported by the U.S. Department of Energy under Contract 
No. DE-AC03-76SF00098, the BMBF, GSI, and DFG.

\end{document}